%% file: main.tex
\documentclass[letterpaper, 10 pt, conference]{ieeeconf}
\IEEEoverridecommandlockouts                

\usepackage[noformatting,nogeometry]{stdcommands}
\usepackage[]{footmisc}
\usepackage{graphicx}
\usepackage{dsfont} 
\usepackage{amsthm}
\usepackage{amsmath}
\usepackage{amssymb}
\usepackage{url}
\usepackage[utf8]{inputenc} 
\usepackage[ruled,linesnumbered]{algorithm2e}
\usepackage{multirow} 
\usepackage{scalerel}
\usepackage[noend]{algpseudocode}
\usepackage[english]{babel}
\usepackage{mathrsfs}
\usepackage{bm}
\usepackage{hhline}
\usepackage{color}
\usepackage{diagbox}

\usepackage{placeins}

\title{Emergency Vehicles Audio Detection and Localization in Autonomous Driving}
\author{%
    \parbox{\linewidth}{\centering
        Hongyi Sun,
        Xinyi Liu,
        Kecheng Xu,
        Jinghao Miao,
        Qi Luo$^*$
    }%
    \thanks{$^*$ Corresponding author}
    \thanks{All Authors are with Apollo Autonomous Driving USA, a Baidu company, 1195 Bordeaux Dr, Sunnyvale, CA 94089 {\tt\small luoqi06@baidu.com}}%
}

\begin{document}
\maketitle
\thispagestyle{empty}
\pagestyle{empty}

\begin{abstract}

Emergency vehicles in service have right-of-way over all other vehicles. Hence, all other vehicles are supposed to take proper actions to yield emergency vehicles with active sirens. As this task requires the cooperation between ears and eyes for human drivers, it also needs audio detection as a supplement to vision-based algorithms for fully autonomous driving vehicles. In urban driving scenarios, we need to know both the existence of emergency vehicles and their relative positions to us to decide the proper actions. We present a novel system from collecting the real-world siren data to the deployment of models using only two cost-efficient microphones. We are able to achieve promising performance for each task separately, especially within the crucial 10m to 50m distance range to react (the size of our ego vehicle is around 5m in length and 2m in width). The recall rate to determine the existence of sirens is 99.16\% , the median and mean angle absolute error is 9.64° and 19.18° respectively, and the median and mean distance absolute error of 9.30m and 10.58m respectively within that range. We also benchmark various machine learning approaches that can determine the siren existence and sound source localization which includes direction and distance simultaneously within 50ms of latency.
\end{abstract}
\IEEEpeerreviewmaketitle

\input{sections/1-Introduction}
\input{sections/2-Framework}
\input{sections/3-Experiments}

\input{sections/5-Conclusion}

\bibliographystyle{IEEEtran}
\bibliography{IEEEabrv,./refs}

\end{document}

%% file: sections/1-Introduction.tex
\section{Introduction}

\subsection{Motivation}
\label{subsection:motivation}
 We have entered an era where fully autonomous driving technology is evolving at a rapid pace. Over the past three years, a record amount of investment flows into the autonomous driving industry. More than 250 tech companies are competing on launching autonomous driving products and services, including robotaxies, robobuses, and self-driving trucks. A handful of automobile giants have already announced their plans to hit the market with L3 autonomy this year. \cite{greyb_2021} As a result, there is a surging demand for onboard software modules that empower and improve vehicle automation.

Knowing to yield emergency vehicles (EV) on public roads is one of the vital qualities for a fully autonomous vehicle (AV), especially that at L4 and beyond. An AV is expected to take appropriate actions even before an emergency vehicle with an active siren reaches its adjacency. The distance can be larger than the range of object identification by a vision-based perception. More importantly, there is a high chance that vision sensors of AVs are blocked by other vehicles, buildings or trees in urban driving scenarios. Hence, audio detection for EV acts as a crucial role to complement vision-based object detection algorithms.

Additionally, the cost of hardware installation required for audio-based detection is reasonably low compared to the benefits it can provide. In our experiments, we use the audio data collected by general-purpose microphones that are economical yet still obtain satisfactory performance. Therefore, audio-based emergency vehicle detection is worth more exploration, while the literature in this area is relatively limited.

\subsection{Related Work}
\label{subsection:related_work}
Siren detection and sound localization are separate topics in most of the previous and ongoing research.

\subsubsection{Siren Detection}
\label{subsection:siren-detection}
Some research focuses on the detection of EV siren sound based on machine learning or deep learning techniques. \cite{beritelli2006} uses waveform and hand-crafted features to train a binary classifier. \cite{meucci2008} extracts the spectrum of the audio signal in their SVM-based siren detection model training. \cite{carmel2017} uses the module difference function (MDF) then applies fully-connected layers for siren classification. \cite{tran2020acoustic} combines both time-domain features and frequency-domain features, and applies convolution neural networks to encode the signal history. 

Even though these models achieve good performance on siren existence detection, they are not designed with the capability to detect the location of the sound source which is critical for an AV to determine appropriate actions.

\subsubsection{Sound Source Localization}
\label{subsection:sound-source-localization}
\cite{grondin2018lightweight}\cite{lee2020gccphat}\cite{shabtai2019detecting} apply rule-based algorithms based on the time difference of arrivals. These methods are only capable to determine to the direction of the sound source, but lack the capability of distance estimation. Plus, the performance is easily influenced by surrounding noise. Because of the inevitably large ambient noise in urban driving, they are not suitable for AVs. \cite{shabtai2019detecting} only uses the siren data collected by their AVs when the AVs are parked still in the hospital without introducing much wind noise. \cite{juan2021}\cite{yalta2017}\cite{christopher2021} apply learning-based models. However, they focus on locating sounds in a relatively quiet indoor environment and the sound sources are close to the microphone devices within 10m distance range.

\subsubsection{Sequence Models}
\label{subsection:sequence-model}
Sequence models take sequential data as input or output. Sequential data, such as financial data, text and audio, are the data where the order matters. While Convolutional Neural Network (CNN) is mainly developed for computer vision applications, its ability to extract the local features from noisy data has attracted researchers and professionals to its applications with sequential data \cite{tran2020acoustic}\cite{yan2019application}\cite{tay2021pre}. On the other hand, the local connectivity also comes with the limitation in capturing the global interaction. Therefore, transformers that directly handle the global information with the self-attention mechanism 
are gaining popularity in sequence models \cite{vaswani2017attention}\cite{zhang2021dive}. The downside of such a structure is the high memory cost, which is quadratic of the length of the sequence $L$. 

As a result, modifications to transformers are proposed to reduce the space complexity in \cite{li2019enhancing}\cite{wu2021autoformer}\cite{kitaev2020reformer}\cite{zhou2021informer}. LogSparse Transformer proposed by \cite{li2019enhancing} only attends to the previous tokens whose indexes' differences are powers of 2 with a combination of convolutions to enhance its locality-agnostics. Such modification reduces the space complexity to $O(L(logL)^2)$, while limiting its capacity in exploiting the long-term dependencies. Reformer proposed by \cite{kitaev2020reformer} utilizes locality-sensitive hashing and reversible transformer to further reduce the space complexity to $O(LlogL)$. It also uses identical queries and keys to reduce the number of parameters required for training. This could improve the model generalization, but potentially impede the model's expressivity. 

\subsection{Contributions}
\label{subsection:main_contributions}
Our proposed method is based on the backbone in \cite{tran2020acoustic}. Our main contributions are the following:
\begin{enumerate}
\item \textbf{Joint Model Training}: Our model is designed to perform three tasks simultaneously, including siren existence, siren sound source direction, and siren sound source distance detection. As shown in Figure \ref{fig:onboard-arc}, these three tasks share a unified backbone which consists of two streams that process the raw waveform and hand-craft features separately. For each task, we attach a block of fully-connected layers to the backbone and output the result accordingly. The latency to produce all three results online is approximately 50ms.


\item \textbf{Low Cost Devices}: To achieve the goal of siren sound source localization, we only use two microphone devices shown in Figure \ref{fig:integration}. Instead of using more expensive customized microphones, we employ more economical general-purpose devices. The price of each microphone device is less than 100 dollars.

\item \textbf{Auto-labeling Method}: We collect the siren data in real driving environments where the AVs with microphones and emergency vehicles with sirens are all moving at normal speed. The large number of data needed for the deep learning makes it even more difficult to manually label the data in such scenario. Therefore, we develop an auto-labeling method by taking advantage of the centimeter-level accurate localization module installed on Apollo autonomous driving system \cite{apollo_open}.

\item \textbf{Long-Distance Real-World data}: While most of the papers tackling audio challenges use synthetic data, we collect the audio data in real urban driving environments. The signal-noise ratio is more stable and the noise is more stationary in the synthetic data. Data collected by holding microphones still can also result in a "reality gap" to the real autonomous driving situation, as the wind noise caused by the moving speed is a significant factor for model performance. In our data set, the distance and the ambient noise during each audio sample can vary in a wide spectrum. The sound sources can be as far as 100m in distances, so the trained models can be generalized to actual autonomous driving testing and operation.

\end{enumerate}

%% file: sections/2-Framework.tex
\section{Framework}
 
\subsection{Hardware Integration}
\label{sec:hardware-integration}
\begin{figure}[!htb]
\centering
  \includegraphics[width=0.4\textwidth]{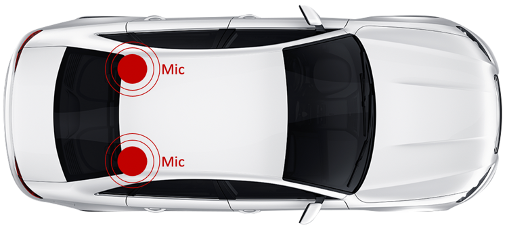}
  \caption{Microphone devices integration.}
  \label{fig:integration}
\end{figure}

Two microphone devices are installed at the rear top of the autonomous driving vehicle (AV) as shown in Figure \ref{fig:integration} after a trade-off between the sound quality and wind noise caused by the high-speed movement of AVs. Each microphone has 4 channels located at the front, back, left and right. There are 8 channels in total. These microphones are connected to the onboard computing unit through USB interfaces.

\subsection{Onboard Architecture}
In this section, we introduce the onboard software architecture of the siren detection module on Apollo autonomous driving platform as shown in Figure \ref{fig:onboard-arc}. It supports multiple-channel audio signal inputs, signal pre-processing and multi-functional model inference.

\subsubsection{Signal Input}
The signal input is collected from the 8 microphone channels as described in Section \ref{sec:hardware-integration}. It takes 1.5 seconds of audio signals at 48k sample rate from every channel. 

\subsubsection{Signal Pre-processing}
The pre-processing step applies a band-pass filter to filter out noise in the frequency range where the sound is less likely to be siren. This processed data are then dumped into the database for model training. The details will be described in Section \ref{sec:offboard-architecture}.

\subsubsection{Model Inference}
\begin{figure*}[!htb]
  \includegraphics[width=0.8\textwidth]{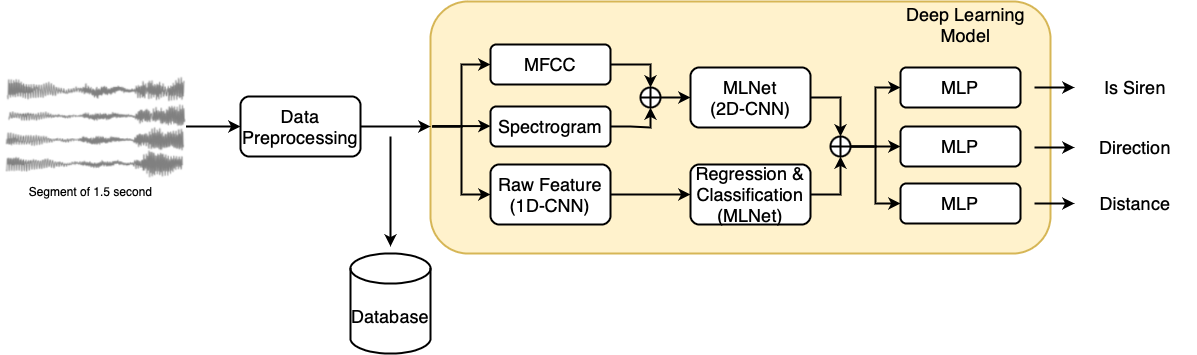}
  \centering
  \caption{Onboard architecture of the siren detection module on Apollo autonomous driving open-source platform.}
  \label{fig:onboard-arc}
\end{figure*}

The model inference part takes the pre-processed signals from these 8 channels, and then applies a neural network model. The output consists of three parts: 1) if siren sound exists 2) siren sound source direction, and 3) siren sound source distance to the autonomous driving vehicle (AV). The details of the model structure can be found in Figure \ref{fig:onboard-arc}.

\subsection{Offboard Architecture}
\label{sec:offboard-architecture}
\begin{figure}[!htb]
  \includegraphics[width=0.48\textwidth]{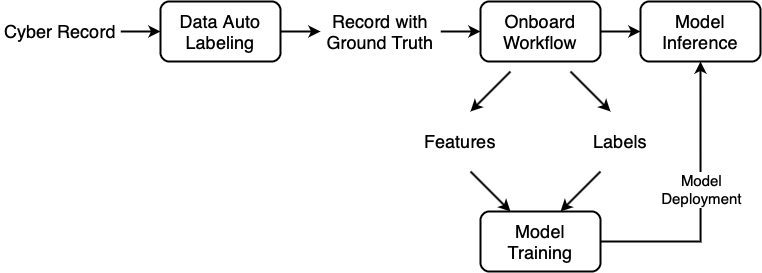}
  \caption{Offboard architecture of the siren detection module on Apollo autonomous driving open-source platform.}
  \label{fig:offboard-arc}
\end{figure}

The offboard architecture is described in Figure \ref{fig:offboard-arc}, which includes data auto labeling, feature extraction and model training.
\subsubsection{Data Auto Labeling}
\label{sec:data-auto-labeling}
One of the straightforward ways to collecting siren data is to follow an emergency vehicle with the siren on, and record the sound from different angles and distances. However, it is impossible (and possibly illegal) to measure the real-time positions of emergency vehicles and it is even harder for a person to recognize or manually label it offline. Deep learning models need hundreds of thousands of data points, therefore manual labeling is not feasible. 

We design an automatic labeling algorithm to label is-siren binary classes, sound source direction and distance relative to the AV. A siren device is installed on another vehicle with Apollo open-source autonomous driving system that has a centimeter-level precise localization module. The locations of vehicles are recorded and later serve as ground truth labels. Vehicles are moving during our data collection, and their locations during the time frame for each data point are changing at different rates. We use the locations of the vehicles at the end of each data point to calculate the direction angle and the distance of the sound source relative to our AVs. The location message at the time of our interest can be either extract from the localization module or calculated by linear interpolation of locations from two nearest timestamps.

All siren data with distances above 100m are discarded because the siren sound is hardly perceivable in that range.

\subsubsection{Feature Extraction}
\label{sec:feature-extraction}
For the purpose of the data consistency between onboard and offboard, we use the data pre-processing part of the onboard workflow to dump the features needed in the model training. The extracted features consist of 8 channels of 1.5 seconds audio signals after the band-pass filter.

\subsubsection{Model Structure}
\label{sec:model-structure}
We deploy the SirenNet model proposed by Tran and Tsai \cite{tran2020acoustic} as our base model. There are two streams of the model to process two kinds of features. One stream processes raw waveform and the other processes the combination of Mel-frequency cepstral coefficients (MFCC) and log-mel spectrogram. These two streams work as the backbone of our model. 

We improve the base model by doing the following upgrades:

\paragraph{Multi-Function MLP}
The representations of these features given by the two streams are concatenated to feed into multilayer perceptrons (MLP). While the SirenNet has only one final output that denotes the possibility of the audio input to be siren sound, our model generates more outputs for siren sound source localization: the direction angle and the distance to uniquely determine the position of the sound source. For the direction angle label, we do not use a single angle value on a scale of (0, 2$\pi$) directly, as the model itself cannot tell 0 and 2$\pi$ are the same. We use sine and cosine values of the direction angle for it to learn. Each MLP generates the output(s) for each task separately as shown in Figure \ref{fig:onboard-arc}.

\paragraph{Attention Mechanism}
In addition to CNN deployed in the base model, we also experiment with transformer and reformer mentioned in \ref{subsection:sequence-model} to process the features from the waveform. To overcome the memory constraint from the length (0.5s * 48k * 8) of each audio data point, we consider 1000 consecutive data points as one token in our sequence data. Since the self-attention mechanism ignores the order of the input data, we add the positional encoding as suggested by \cite{vaswani2017attention}. Instead of summing the positional encoding with the audio encoding, we concatenate them together by setting the length of positional encoding as 100. Note that we also experiment with the summing method but find that the concatenation method works consistently better with our data.

%% file: sections/3-Experiments.tex
\section{Experiments and Results}
\label{sec:experiment_and_results}

\subsection{Data Collection}
\label{sec:data_collection}
\begin{figure}
  \centering
  \includegraphics[width=4cm, height=4cm]{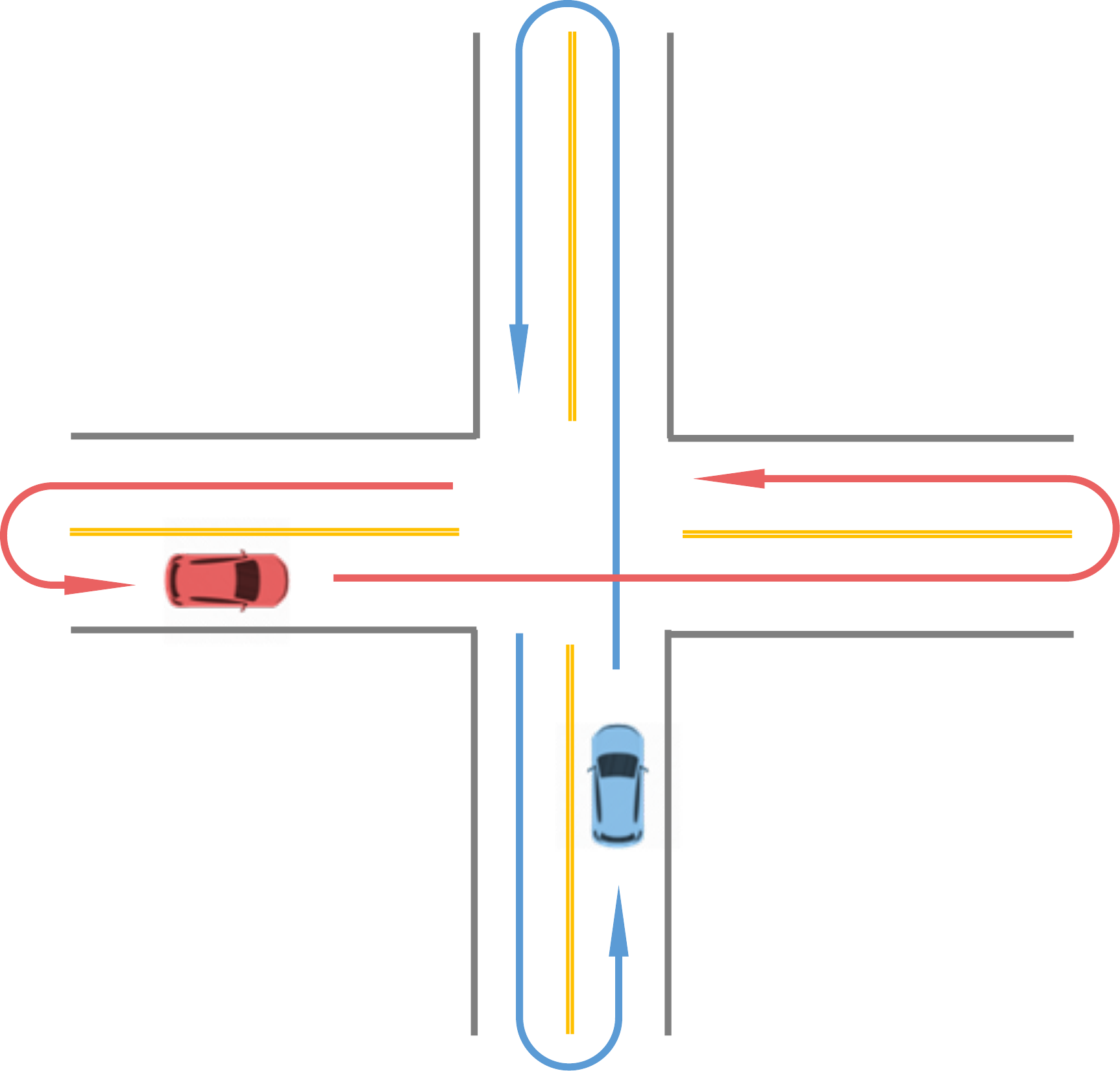}\hspace{0.6cm}
  \includegraphics[width=3cm, height=4cm]{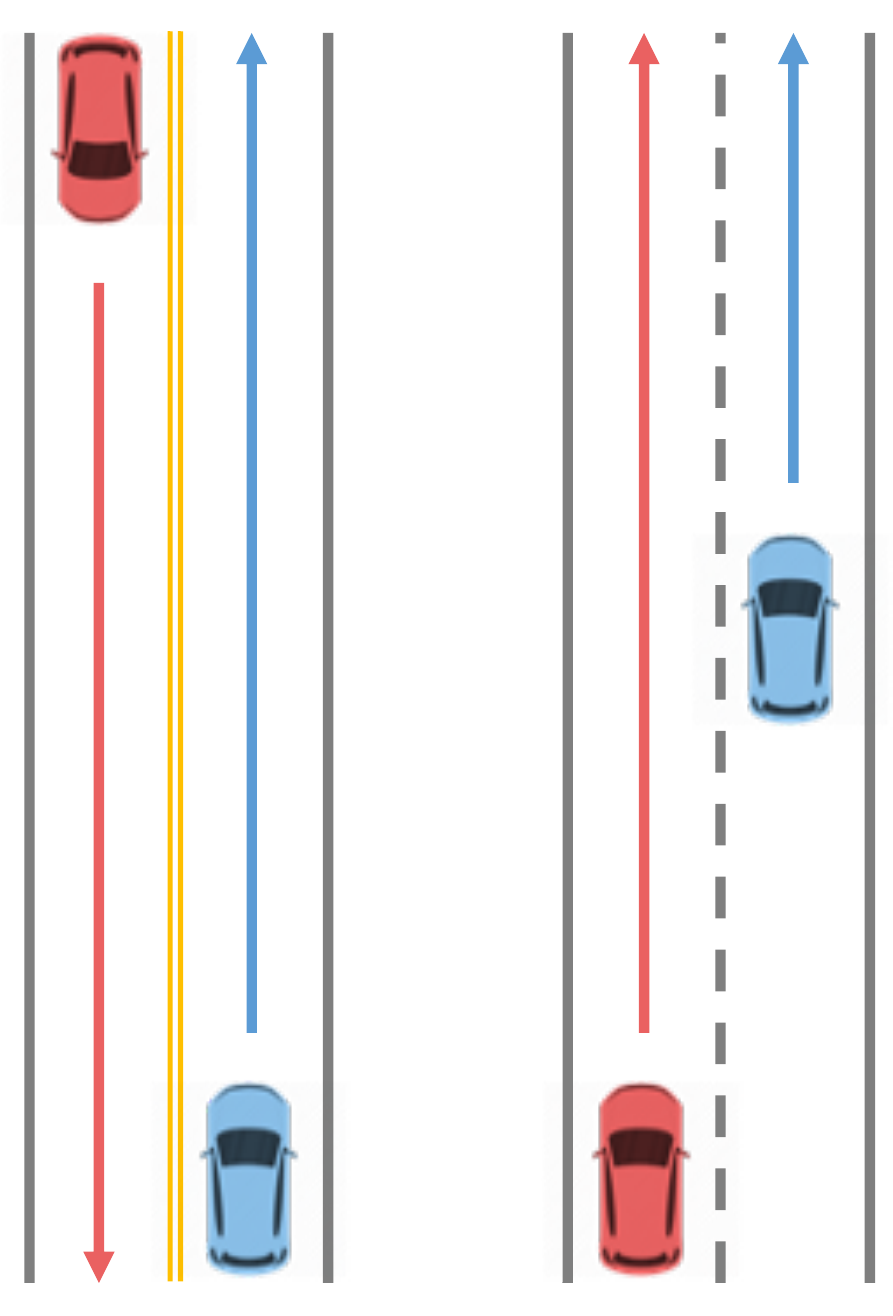}
  \caption{Three data collection scenarios, where the blue one is the ego vehicle, the red one is the emergency vehicle. The left part is an intersection scenario where the two vehicles' trajectories are orthogonal to each other. The middle and right parts are the scenarios of regular road where the two vehicles' trajectories are parallel. The two vehicles' are in opposite directions in the middle part and in the same direction in the right part.}
  \label{fig:data_scenario}
\end{figure}

We set up three scenarios for data collection as shown by Figure \ref{fig:data_scenario}. Under these settings, we are able to collect siren sound data from different directions at different distances with different ambient noise levels. Those collected records, including the siren sound recorded from the microphones and also the localization information of vehicles, are uploaded to our cloud platform. The auto labeling tool \ref{sec:data-auto-labeling} is running in the cloud to generate machine learning-ready signal features and ground truth labels.

We collect a total of 20 hours of positive records with siren sound and 10 hours of negative records without siren sound in urban driving environments. Then we down-sample some positive data where the emergency vehicle with the siren on is strictly in front of and behind our AV to balance the direction distribution of our data. When data sets are split into training, validation, and testing set, we make sure that the data from the training set and the data from the valid/test set are not collected on the same day because the training data consists of consecutive sliding time windows of 1.5s interval of data. The model is validated and tested on data sets collected from different days from those in training. In this way, we decorrelate the data for better generalization. The samples from training, validation, and testing roughly follow a ratio of 8:1:1.

\subsection{Experiment Setup}
\label{sec:experiment_setup}

\subsubsection{Data Preparation}
\label{sec:data_preparation}
The audio data is recorded at 48 kHz sampling rate by two 4-channel microphones. Every 1.5s length of audio is considered as one data point and we collect data points every 0.17s. We experiment with a time window interval of 1.5s, 1s, 0.5s and 0.25s as model input and find that 0.5s input achieves the best overall performance so we report all the results with 0.5s input. Band-pass filter is used to only keep the signal with frequency from 500Hz to 1800Hz so the model can focus on the part of the audio that is more likely to be a siren sound.

We consider the direction of the siren sound as 0 when it is right in front of our ego vehicle. When the siren sound is at the left side of the ego vehicle, the direction angle is positive; when the siren sound is at the right side of the ego vehicle, the direction angle is negative. To ensure the sum of the square of the sine and cosine values of direction angle is one, we normalize these two outputs by dividing them with the square root of their sum of the square. The two normalized values are then used to calculate the direction angle output $\hat{\theta}$. When the absolute difference between the direction angle output $\hat{\theta}$ and the ground truth angle $\theta$ is larger than $\pi$, we use $2\pi - \hat{\theta} + \theta$ in calculating the mean absolute error for the report purpose.

\subsubsection{Model Training}
\label{sec:model_training}
We conduct our experiments on a server that runs the Ubuntu 18.04 operating system with the models implemented with 1.5.1 PyTorch framework. Adam is used as the optimizer and the initial learning rate is set as $10^{-5}$. The minimum learning rate is set as $10^{-9}$. 

We use the binary cross entropy loss for is-siren binary classification. The sum of squared error is used for the sine and cosine of the direction angle and the distance detection. Since the loss calculated for our three tasks are at different magnitudes, the final loss of the whole model for the gradient computation is the weighted sum of these three tasks. The weights are subject to our experiments.

The accuracies are reported for the is-siren binary classification task. Mean absolute errors are calculated for both the direction angle and distance detection tasks. We train the model to detect the direction angle and distance of siren sound by masking the negative input data in calculating the loss for these two tasks. We choose the models with the best performance on the valid set and report their performance on the test set.

\subsection{Performance Study}
\label{sec:performance_study}

\begin{table*}
\caption{Performance Comparison between CNN, Transformer and Reformer models}
\label{Tab: performance-comparison}
\centering
\begin{tabular}{|c|c|c|c|c|} 
 \hline 
 Model & Weights & Is\_siren (Accuracy) & Angle Error (°) & Distance(m) \\ 
 \hline \hline
 \multirow{6}{*}{CNN} & (1,0,0) & 0.944 & NA & NA \\
 & (0,1,0) & NA & 34.60° & NA \\
 & (0,0,1) & NA & NA & 14.226 \\
 & (10,10,0.008) & 0.896 & 37.13° & 15.221 \\
 & (10,5,0.008) & 0.887 & 38.33° & 14.761 \\
 & (10,5,0.005) & 0.796 & 40.28° & 14.432 \\
 \hline
 \multirow{6}{*}{Transformer} & (1, 0, 0) & 0.940 & NA & NA \\
 & (0, 1, 0) & NA & 46.12° & NA \\
 & (0, 0, 1) & NA & NA & 13.560 \\
 & (10, 10, 0.002) & 0.870 & 46.87° & 15.335 \\
 & (10, 10, 0.008) & 0.832 & 50.25° & 14.318 \\
 & (10, 5, 0.005) & 0.839 & 51.97° & 14.357 \\
 \hline
  \multirow{6}{*}{Reformer} & (1, 0, 0) & 0.938 & NA & NA \\
 & (0, 1, 0) & NA & 47.96° & NA \\
 & (0, 0, 1) & NA & NA & 13.606 \\
 & (10, 10, 0.002) & 0.862 & 47.78° & 14.812 \\
 & (10, 10, 0.008) & 0.834 & 51.45° & 14.763 \\
 & (10, 5, 0.005) & 0.856 & 50.82° & 14.385 \\
\hline
\end{tabular}
\end{table*}

Table~\ref{Tab: performance-comparison} shows that CNN models have better performance than the others in general. All models show comparable performances when it comes to the siren existence detection task. For the distance prediction task, the CNN model performs moderately worse than either the Transformer or the Reformer model. However, the CNN model performs significantly better than the Transformer and the Reformer model in siren direction angle prediction. The angle MAE of the CNN model is 34.6°, which is 25\% better than the angle MAE, 46.1°, of the Transformer model and 29\% better than the angle MAE, 48.0°, of the Reformer model. 

The key factor in detecting the direction angle of the sound source is the difference time of arrivals among different channels, especially among different microphones for significant time differences. CNN is able to compute the relationship of the signals from different channels in each batch. Transformer and Reformer should be able to do that theoretically. However, they rely on positional encoding to understand the timestamp of the signals in each token input. This information requires more training of the model to capture aside from the signal information, thus requiring a much larger data set. We speculate that 264,506 positive training data points after data pre-processing result in the underfitting of Transformer-based models.

Table~\ref{Tab:breakdown-performance} shows the performance breakdown of CNN models according to the distance ranges in which the siren is from the AV. Since there is no distance label for a nonexistent siren, there is only the recall rate change of our model according to the distance between the siren sound source and our AV. When the sound source is less than 20m away from the AV, the recall rate is as high as 100\%. When the sound source is 40m away from the ego vehicle, the recall rate can still remain over 99\%. The recall rate is monotonically decreasing with the increase of the distance. The frequency of a siren sound is decreasing with increasing distance according to the Doppler effect, which can potentially lead to the overlapping of the siren sound frequency and noise frequency in urban driving. On the other hand, the siren sound intensity also decreases with the distance increases, thus resulting in a low signal-noise ratio.

\begin{figure}[htb]
  \includegraphics[width=0.48\textwidth]{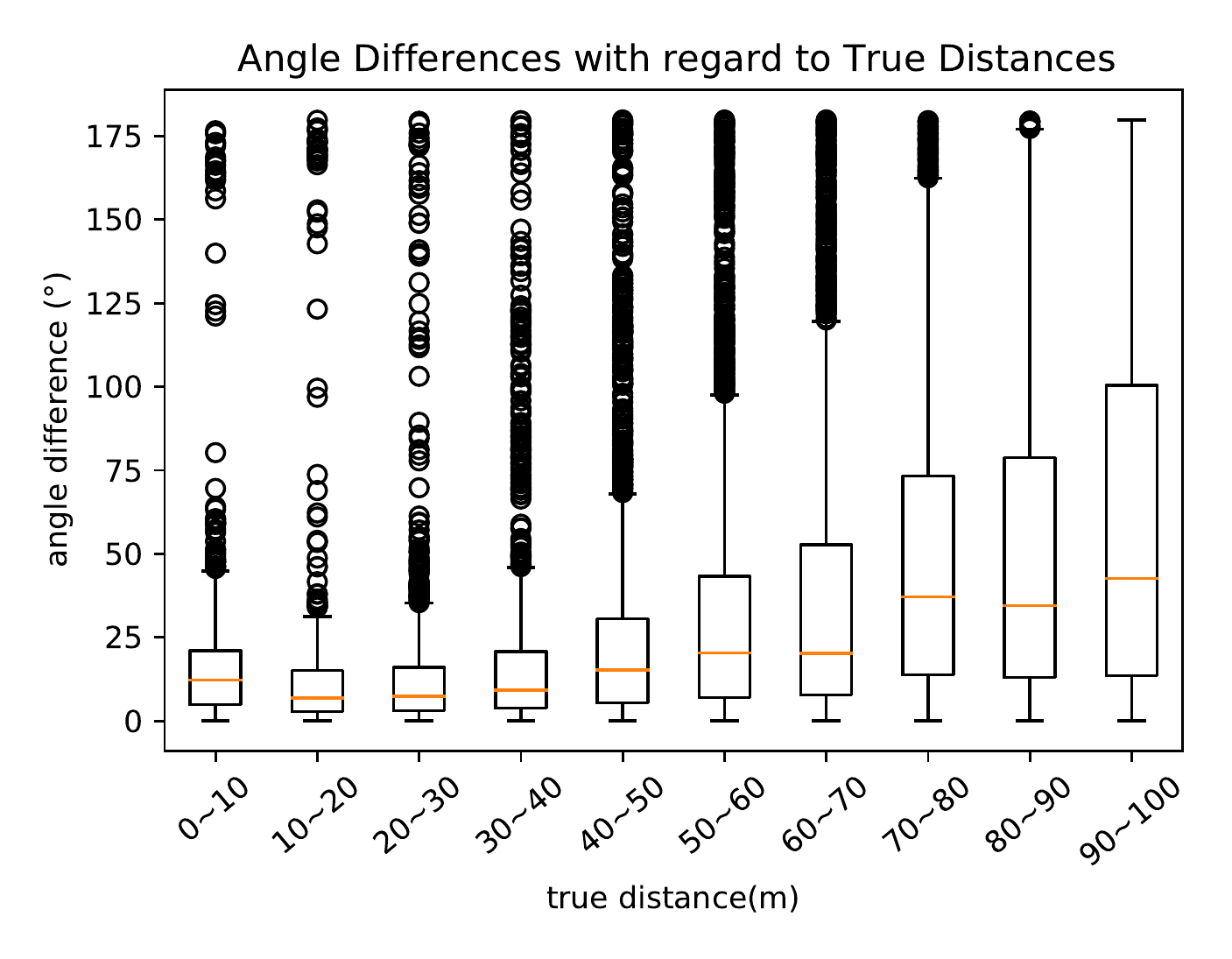}
  \caption{CNN angle error analysis.}
  \label{fig:angle-error}
\end{figure}

In contrast, the angle MAE is decreasing with the increase of the distance when the sound source is less than 30m away from the AV. This is because our data is collected when both the sound source and the microphones are moving. Even though we only input 0.5s audio sample to the model, the relative angle of the siren to our ego vehicle can change dramatically during this time when the siren is close to the ego vehicle. As shown in Figure \ref{fig:angle-error}, the median of absolute angle error is only around 7° when the distance is from 10m to 30m. When the distance is larger than 30m, the angle MAE is increasing as the distance is increasing.

\begin{figure}[!htb]
  \includegraphics[width=0.48\textwidth]{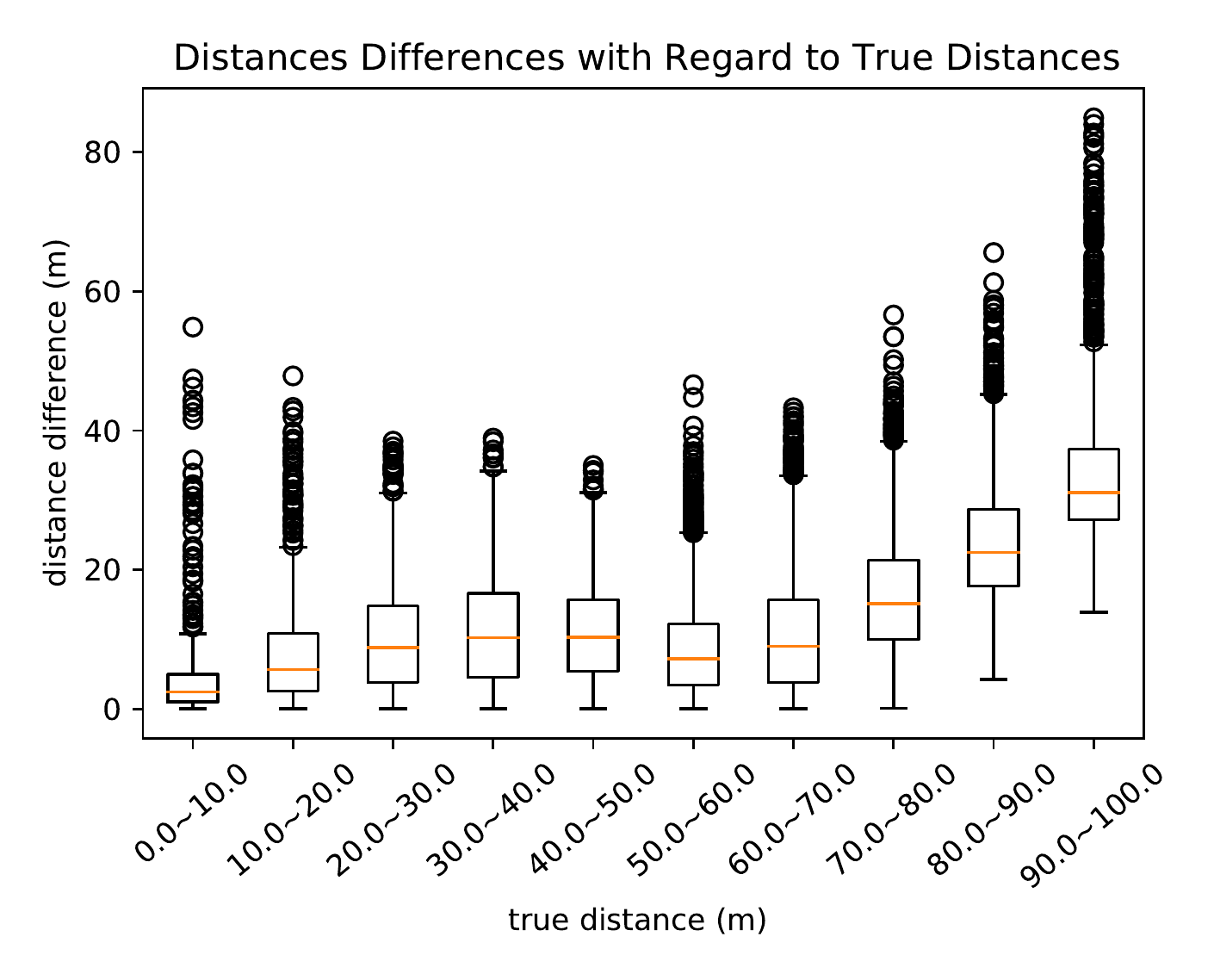}
  \caption{CNN distance error analysis.}
  \label{fig:distance-error}
\end{figure}

The distance detection MAE has an increasing trend with increasing distance overall, also shown by Figure \ref{fig:distance-error}. However, when the distance is between 20m to 70m, the distance MAE is relatively stable. When the distance is larger than 70m, the MAE of distance detection increases rapidly as the distance increases. This could be the result of the quality constraint of the microphones as we install general-purpose microphones instead of customized microphones that are more costly.

\begin{table*}[!htb]
\caption{Performance Breakdown}
\label{Tab:breakdown-performance}
\centering
\begin{tabular}{|c|c|c|c|c|c|c|c|c|c|c|} 
 \hline 
\diagbox{Metric}{Distance Range(m)} & 0-10 & 10-20 & 20-30 & 30-40 & 40-50 & 50-60 & 60-70 & 70-80 & 80-90 & 90-100 \\ 
 \hline
 Recall (\%) & 100 & 100 & 99.64 & 99.15 & 98.49 & 95.78 & 95.95 & 96.99 & 95.37 & 95.21 \\
 \hline
Angle MAE (°) & 23.54 & 15.39 & 13.56 & 17.55 & 26.30 & 32.98 & 35.59 & 50.36 & 51.83 & 59.14 \\
 \hline 
Distance MAE (m) & 5.52 & 8.16 & 10.15 & 11.26 & 11.09 & 8.83 & 10.74 & 16.29 & 23.98 & 33.98 \\
 \hline
\end{tabular}
\end{table*}

Table~\ref{Tab: performance-comparison} also shows the performance of different models that are trained to produce the outputs for three different tasks concurrently. We choose the best model from CNN, Transformer and Reformer and fine-tune the last MLP layers associated with different tasks. The performance of these models after fine-tuning is reported in Table~\ref{Tab:fine-tuning}. The model with the best performance is still a CNN model. The accuracy for this model is 89.4\% while the angle MAE is 36.6° and the distance MAE is 14.2m.

\begin{table*}[!htb]
\caption{Performance Comparison between CNN, Transformer and Reformer models}
\label{Tab:fine-tuning}
\centering
\begin{tabular}{|c|c|c|c|c|} 
 \hline 
 Model & Weights & Is\_siren (Accuracy) & Angle Error ($\pi$) & Distance(m) \\ 
 \hline \hline
 CNN & (10, 10, 0.008) & 0.894 & 0.639(36.61°) & 14.184 \\
 \hline
Transformer & (10, 10, 0.002) & 0.880 & 0.817(46.81°) & 13.351 \\
 \hline
Reformer & (10, 10, 0.002) & 0.869 & 0.836(47.90°) & 13.786 \\
\hline
\end{tabular}
\end{table*}


%% file: sections/5-Conclusion.tex
\section{Conclusions}

 Our data collection system can collect the real-world sirens data in urban driving environments without tedious human labor. With cost-effective microphones, we are able to achieve excellent performance in three tasks when emergency vehicles are within a distance that is crucial for AV to react, 10m to 50, by training the model with the siren data within 100m distance range. This emergency audio detection system compensates for the limitation of vision-based object detection with state-of-the-art performance. We also test our system in actual autonomous driving scenarios and receive satisfactory feedback. The robustness of the model performance degrades when the siren distance is beyond a certain distance, i.e. 70m in our model. This can be potentially improved by utilizing a larger number of microphone arrays and more data collected.